\documentstyle[psfig]{aipproc}

\begin{document}

\hspace*{11cm} JLAB-THY-00-35

\hspace*{11cm} ADP-00-50/T430

\title{Quark-Hadron Duality \\ in Electron Scattering\footnote{
Talk presented at the Second Workshop on Physics with an
Electron Polarized Light-Ion Collider (EPIC), MIT, Sep.14-16, 2000.}}

\author{W. Melnitchouk}
\address{Jefferson Lab, 12000 Jefferson Avenue,
	Newport News, VA 23606, and					\\
	Special Research Centre for the Subatomic Structure of Matter,	\\
	Adelaide University, Adelaide 5005, Australia}

\maketitle

\begin{abstract}
Quark-hadron duality addresses some of the most fundamental issues
in strong interaction physics, in particular the nature of the
transition from the perturbative to non-perturbative regions of QCD.
I summarize recent developments in quark-hadron duality in
lepton--hadron scattering, and outline how duality can be studied
at future high-luminosity facilities such as Jefferson Lab at 12~GeV
or an electron--hadron collider such as EPIC.
\end{abstract}

\section{Introduction}

Understanding the structure and interaction of hadrons in terms of the
quark and gluon degrees of freedom of QCD is the greatest unsolved
problem of the Standard Model of nuclear and particle physics. 
If one accepts QCD as the correct theory of the strong interactions,
then the transition from quark-gluon to hadron degrees of freedom should
in principle amount to a change of basis, with all physical quantities
independent of which basis is used.
However, although the duality between quark and hadron descriptions is
formally exact, in practice the necessity of truncating any Fock state
expansion means that the extent to which duality holds reflects the
validity of the truncations under different kinematical conditions and
in different physical processes.
Quark-hadron duality is therefore an expression of the relationship
between confinement and asymptotic freedom, and is intimately related to
the nature of the transition from non-perturbative to perturbative QCD.

In nature, the phenomenon of duality is in fact quite general and can be
studied in a variety of processes, such as $e^+ e^- \rightarrow$ hadrons,
or heavy quark decays \cite{HQ}.
One of the more intriguing examples, initially observed some 30 years
ago, is in inclusive inelastic electron--nucleon scattering.

\section{Bloom-Gilman Duality}

In studying inelastic electron scattering in the resonance region and
the onset of scaling behavior, Bloom and Gilman \cite{BG} found that the
inclusive $F_2$ structure function at low $W$ generally follows a global
scaling curve which describes high $W$ data, to which the resonance
structure function averages.
More recently, high precision data on the $F_2$ structure function
from Jefferson Lab \cite{F2JL}, shown in Fig.~1, have confirmed the
earlier observations, demonstrating that duality works remarkably well
for each of the low-lying resonances, including the elastic, to rather
low values of $Q^2$ ($\sim 0.5$~GeV$^2$).

\begin{figure}[h]
\vspace*{6cm}
\includegraphics{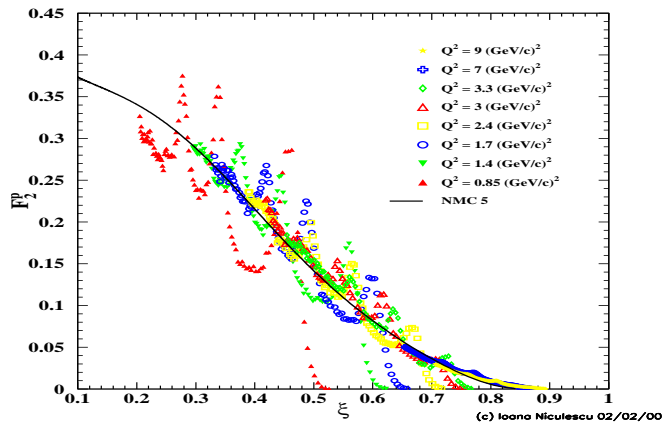}
\caption{Proton $F_2$ structure function in the resonance region
	(data from a compilation \protect\cite{F2JL} of JLab and SLAC 
	data); solid line is a fit to large-$W$ deep-inelastic data at
	$Q^2=5$~GeV$^2$.}
\end{figure}

Before the advent of QCD, Bloom-Gilman duality was initially interpreted
in the context of finite-energy sum rules \cite{DHS}.
Formulated originally for hadron-hadron scattering, they relate the
high-energy behavior of amplitudes, described within Regge theory in
terms of $t$-channel Regge pole exchanges, to the behavior at low energy,
which can be well described by a sum over a few $s$-channel resonances
\cite{VEN}.
Later Harari \cite{HARARI} suggested a generalization of the duality
picture to include both resonant and non-resonant background contributions
to cross sections, in which resonances were dual to non-diffractive Regge
pole exchanges, while the non-resonant background was dual to Pomeron
exchange.
For electron scattering, this translates into a duality between resonances
and valence quarks (whose small $x \sim 1/s$ behavior is given in Regge
theory by non-diffractive Reggeon exchanges), with the background dual to
sea quarks (for which the small-$x$ behavior is determined by diffractive
Pomeron exchange).

In QCD, Bloom-Gilman duality can be reformulated in the language of the
operator product expansion, in which QCD moments of structure functions
are organized according to powers of $1/Q^2$ \cite{RUJ}.
The leading terms are associated with free quark scattering, and are
responsible for the scaling of the structure function, while the $1/Q^2$
terms involve interactions between quarks and gluons and hence reflect
elements of confinement dynamics.
The weak $Q^2$ dependence of the low moments of $F_2$ is then interpreted
as indicating that the non-leading, $1/Q^2$-suppressed, interaction terms
do not play a major role even at low $Q^2$ ($\approx 1$~GeV$^2$).

An important consequence of duality is that the strict distinction between
the resonance and deep-inelastic regions is quite artificial.
As observed by Ji and Unrau \cite{JU}, at $Q^2 = 1$~GeV$^2$ around 70\% of
the total cross section comes from the resonance region,
$W < W_{\rm res} = 2$~GeV, however, the resonances and the deep-inelastic
continuum conspire to produce only about a 10\% correction to the lowest
moment of the scaling $F_2$ structure function at the same $Q^2$.
The deep-inelastic and resonance regions are therefore intimately related,
and properly averaged resonance data can help us understand the
deep-inelastic region \cite{ISGURTALK,JIMV}.
This has immediate implications for global analyses of parton distribution
functions, in which the standard procedure is to omit from the data base
the entire resonance region below $W = 2$~GeV.
This is of practical relevance especially for the large-$x$ region, where
deep-inelastic data are scarce \cite{LARGEX}.

\section{Testing the Bounds of Duality}

Since the details of quark--hadron duality are process dependent,
there is no reason to expect the accuracy to which it holds and the
kinematic regime where it applies to be similar for different
observables.  
In fact, there could be qualitative differences between the workings
of duality in spin-dependent structure functions and spin-averaged
ones \cite{CM,JM}, or for different hadrons --- protons compared with
neutrons, for instance.

At present there are data on the $F_2$ structure function of the proton
and deuteron \cite{F2JL}, but little or no information at all exists on
the spin-dependent $g_1$ and $g_2$ structure functions (which correspond
to differences of cross sections), nor on the longitudinal to transverse
structure function ration, $R$.
It is vital for our understanding of duality and its practical
exploitation that the spin and flavor dependence of duality,
as well as its nuclear dependence, be established empirically.

Another largely unexplored domain with potentially broad applications is
the production of mesons ($M$) in semi-inclusive electron scattering,
$e N \rightarrow e' M X$.
At high energy the scattering and production mechanisms factorize, with
the cross section at leading order in QCD given by a simple product of the
structure function and a quark $\rightarrow$ meson fragmentation function,
as in Fig.~2.
In terms of hadronic variables the same process can be described through
the excitation of nucleon resonances, $N^*$, and their subsequent decays
into mesons and lower lying resonances, $\widetilde N^*$.
The hadronic description is rather elaborate, however, as the production
of a fast outgoing meson in the current fragmentation region at high
energy requires non-trivial cancellations of the angular distributions
from various decay channels \cite{ISGURTALK,JIMV}.
Heuristically, the duality between the quark and hadron descriptions
of semi-inclusive meson production (see Fig.~2) can be written as:
\begin{eqnarray}
\sum_{N^*,\widetilde N^*}
F_{\gamma^* N \to N^*}(Q^2,W^2)\
{\cal D}_{N^* \to \widetilde N^* M}(W^2,\widetilde W^2)\
&\sim&\ 
\sum_q e_q^2\ q(x)\ D_{q \to M}(z)\ ,
\nonumber
\end{eqnarray}
where $D_{q \to M}$ is the quark $\to$ meson fragmentation function
for a given $z=E_M/\nu$,
$F_{\gamma^* N \to N^*}$ is the $\gamma^* N \to N^*$ transition form
factor, which depends on the mass of the virtual photon and the excited
nucleon ($W = M_{N^*}$), and
${\cal D}_{N^* \to \widetilde N^* M}$ is a function representing the decay
$N^* \to \widetilde N^* M$, where $\widetilde W$ is the invariant mass of
the final state $\widetilde N^*$.

\begin{figure}[t]
\vspace*{4cm}
\includegraphics{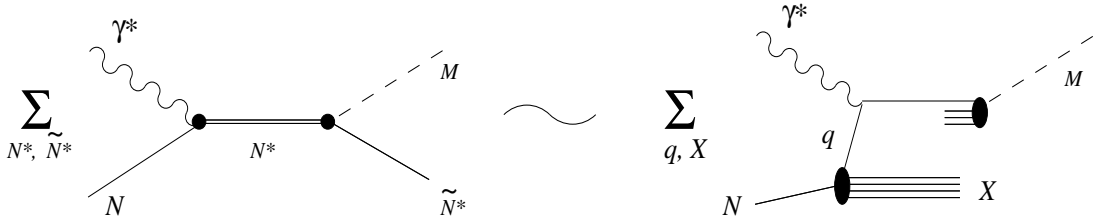}
\caption{Duality between descriptions of semi-inclusive meson production
	in terms of quark (right) and nucleon resonance (left) degrees
	of freedom.}
\end{figure}

The virtue of semi-inclusive production lies in its ability to identify
individual quark species in the nucleon by tagging specific mesons in the
final state, enabling both the flavor and spin of quarks and antiquarks
to be systematically determined.
To what extent factorization applies at lower energy is an open question,
and the signatures of duality in the resonance region of semi-inclusive
scattering are still under investigation.
Confirmation of duality in inclusive hadron production would clearly open
the way to an enormously rich semi-inclusive program in the pre-asymptotic
regime, allowing unprecedented spin and flavor decomposition of quark
distributions.

\section{Conclusion}

Quark-hadron duality offers the prospect of addressing the physics of
the transition from the strong to weak coupling limits of QCD, where
neither perturbative QCD nor effective descriptions such as chiral
perturbation theory are applicable.
While considerable insight into quark-hadron duality has already been
gained from recent theoretical studies, it will be important in future
to understand more quantitatively the features of the electron scattering
data in the resonance region and the phenomenological $N^*$ spectrum in
terms of realistic models of QCD.

On the experimental side, the spin and flavor dependence of duality can be
most readily accessed through semi-inclusive scattering, which requires a
facility with both high luminosity and a high duty factor.
Jefferson Lab at 12~GeV would be an ideal facility to study meson
production in the current fragmentation region at moderate $Q^2$, allowing
the onset of scaling to be tracked in the pre-asymptotic regime.
On the other hand, the higher center of mass energy available at an
electron-hadron collider, such as EPIC, would, despite a lower luminosity,
enable measurement of semi-inclusive cross sections to larger values of
$Q^2$ where perturbative QCD is more readily applicable, and factorization
of the current and target fragmentation regions less problematic.
Furthermore, unlike fixed-target facilities, a collider mode would allow
unique access to hadrons produced in the target fragmentation region.
An understanding of duality for target fragments would be the next
challenge for electron scattering experiments.

\section*{Acknowledgements}

I would like to thank F.E.~Close, R.~Ent, N.~Isgur, S.~Jeschonnek,
C.~Keppel and J.W.~Van~Orden for many informative and stimulating
discussions about duality.
This work was supported by the Australian Research Council,
and U.S. Department of Energy contract DE-AC05-84ER40150.


\end{document}